\documentclass[useAMS,usegraphicx,usenatbib]{mn2e}

\usepackage{ulem}
\usepackage{color}
\usepackage{graphics,graphicx}
\usepackage{times} 
\usepackage{amssymb}
\usepackage{amsmath}
\usepackage{lscape}
\usepackage{url}
\usepackage{multirow}
\usepackage{ctable}
\usepackage{threeparttable}
\newif\ifAMStwofonts
\AMStwofontstrue
\title{Explaining the Hard Excesses in AGN}

\author[D.\,J. Walton et al.]
{\parbox{7.in}{D.\,J. Walton \thanks{E-mail: dwalton@ast.cam.ac.uk},
R.\,C. Reis and A.\,C. Fabian \\ 
\footnotesize
\it{Institute of Astronomy, Cambridge University, Madingley Road, Cambridge, CB3 0HA}}}
\date{}

\def\xmm{{\it XMM-Newton~\/}}

\def\suzaku{{\it SUZAKU}}

\def\chandra{{\it Chandra}}

\def\epicpn{{\it EPIC}{\rm-pn}}
\def\epicmos1{{\it EPIC}{\rm-MOS1~\/}}
\def\epicmos2{{\it EPIC}{\rm-MOS2 ~\/}}
\def\epicmos{{\it EPIC}{\rm-MOS}}
\def\rgs{\rm RGS}
\def\xis{{\rm XIS}}
\def\pin{{\rm PIN}}
\def\hxd{{\rm HXD}}

\def\ks{\hbox{$\rm\thinspace ks$}}

\def\deg{$^{\circ}$}

\def\kmps{\hbox{$\rm\thinspace km~s^{-1}$}}

\def\H0{{\rm ~km~s^{-1}~Mpc^{-1}}}

\def\kev{\hbox{$\rm\thinspace keV$}}

\def\ctps{\hbox{$\rm\thinspace count~s^{-1}$}}

\def\atpcm{{\rm atom~cm$^{-2}$}}

\def\ergpcmsqps{\hbox{$\rm\thinspace erg~cm^{-2}~s^{-1}$}}

\def\ergps{\hbox{erg~s$^{-1}$}}
\def\ergcmps{\hbox{\rm erg~cm~s$^{-1}$}}

\def\chisq{{\chi^{2}}}
\def\rchi{{$\chi^{2}_{\nu}$~\/}}

\def\reflionx{\rm{\small REFLIONX}}

\def\kdblur{\rm{\small KDBLUR}}

\def\pexrav{\rm{\small PEXRAV}}
\def\pexriv{\rm{\small PEXRIV}}

\def\heasoftv{\hbox{\rm{\small HEASOFT~\/}}}
\def\xselect{\hbox{\rm{\small XSELECT~\/}}}

\def\ftool{\hbox{\rm{\small FTOOL}}}

\def\grppha{\hbox{\rm{\small GRPPHA~\/}}}
\def\mathpha{\hbox{\rm{\small MATHPHA}}}
\def\addascaspec{\hbox{\rm{\small ADDASCASPEC~\/}}}

\def\hxddtcor{\hbox{\rm{\small HXDDTCOR}}}
\def\hxdpinxbpi{\hbox{\rm{\small HXDPINXBPI}}}
\def\mgtime{\hbox{\rm{\small MGTIME}}}
\def\addascaspec{\hbox{\rm{\small ADDASCASPEC}}}

\def\grid25{\hbox{\rm{\small GRID25}}}

\def\ovii{\hbox{\rm O{\small VII}~\/}}

\def\fexxv{\hbox{\rm Fe\thinspace{\small XXV}}}
\def\fexxvi{\hbox{\rm Fe\thinspace{\small XXVI}}}
\def\ka{$K\alpha$\thinspace}
\def\kb{$K\beta$\thinspace}
\def\feabund{$A_{Fe}$}

\def\eg{{\it e.g.~\/}}
\def\etc{{\it etc.}}
\def\ie{{\it i.e.~\/}}

\def\la{\mathrel{\hbox{\rlap{\hbox{\lower4pt\hbox{$\sim$}}}{\raise2pt\hbox{$<$}}}}}
\def\ga{\mathrel{\hbox{\rlap{\hbox{\lower4pt\hbox{$\sim$}}}{\raise2pt\hbox{$>$}}}}}

\def\d25{D$_{25}$}
\def\nh{{$N_{H}$}}

\def\.25{0.25 keV\thinspace}

\def\rg{$R_{G}$}
\def\rin{$R_{in}$}
\def\rout{$R_{out}$}
\def\rbr{$R_{br}$}

\def\ngc1365{\rm NGC\thinspace1365}
\def\1h0419{\rm 1H\thinspace0419-577}
\def\pds456{\rm PDS\thinspace456}

\begin{document}
\pagerange{\pageref{firstpage}--\pageref{lastpage}}
\pubyear{2010}
\maketitle
\label{firstpage}

\begin{abstract}
A common model invoked to describe the X-ray spectra of active galaxies includes a
relativistically blurred reflection component, which in some cases can be the dominant contributor to the
received flux. Alternative interpretations are often based around complex absorption,
and to date it has proven difficult to determine between
these two viable models. Recent works on \suzaku\ observations of the active nuclei in \ngc1365, \1h0419 and
\pds456 have found the presence of strong X-ray emission at high ($\sim$10-50 \kev)
energies, referred to as `hard excesses', and it has been claimed this emission cannot
be explained with simple disc reflection models. Here we investigate the high energy
emission in these sources by constructing disc reflection models and show that they can
successfully reproduce the observed spectra. In addition, we
find the behaviour of \ngc1365 and \1h0419 in these observations to be broadly
consistent with previous work on disc reflection interpretations.
\end{abstract}


\section{Introduction}

High quality X-ray spectra of active galactic nuclei (AGN) often display complex
deviations from the powerlaw continuum expected from the intrinsic X-ray source. Since
the launches of the \xmm and \chandra\ observatories in 1999, a wealth of high quality
data on these spectral features in the $\sim$0.5--10.0\kev\ energy range has been
obtained, and depending on the choice of continuum, many of them can be interpreted
as either emission or absorption features, with viable models having been proposed
for both cases.

In the disc reflection/light bending interpretation, the
fraction of the powerlaw continuum emitted in the direction of the accretion disc
is reprocessed into an additional `reflected' emission component. If the accretion
disc extends to the innermost stable circular orbit of the black hole, the strong
gravity present will cause the reflection component to appear smooth and can produce
broadened and skewed emission lines. An extreme example of this behaviour is presented by \cite{FabZog09}.
In this case the primary source of the intrinsic powerlaw continuum is also located in a region
of strong gravity, and gravitational light bending focuses some of the emission from
this component down onto the disc, providing an explanation for the spectral variability
often observed; if the location of the primary source varies in distance from
the central black hole, the fraction of the emitted flux focussed onto the disc, and
hence that observed, also varies (see \eg \citealt{lightbending}). Amidst such variations,
the observed spectrum can be totally dominated by the reflection component, even if
the intrinsic flux of the powerlaw component has not changed.

An alternative interpretation is that the observed spectral features may arise as a
result of complex absorption. The absorbing material may be neutral, partially or almost
fully ionised, and could fully cover the source or only obscure some fraction of it.
In addition, the structure of the absorbing material might be complex leading to
absorption by multiple layers of gas with differing physical circumstances and
dimensions. For a review of the absorption processes relevant to this
interpretation see \cite{Turner09a}. Possible suggestions for the origins of such
gas range from the dusty torus and broad emission line region (BLR) clouds expected
in the standard picture of AGN to outflowing (possibly relativistic in some cases)
disc winds, so the velocity structure of the absorbing material is also important.
Appropriate combinations of absorbers such as these can reproduce the observed
spectra, and spectral variability is interpreted as changes in the properties
of the intervening gas, \eg changes in ionisation and the fraction of the intrinsic
source obscured.

Despite the quality and quantity of the data provided by \xmm and \chandra\ it has
proven very difficult to distinguish between these interpretations for individual
cases, a classic example being the highly variable AGN MCG{\thinspace-6-30-15} (see \citealt{Fabian02MCG}
and \citealt{Miller08}). In July 2005 the \suzaku\ X-ray observatory was launched,
offering simultaneous X-ray observations extending up to $\sim$70\kev, providing
potentially important broadband information against which these models can be tested.
Recently, \suzaku\ observations of \ngc1365, \1h0419 and \pds456 have revealed
strong emission above 10\kev, see \cite{Risaliti09c}, \cite{Turner09b} and \cite{Reeves09}
for the respective sources. In each case, the authors claim that reflection models
cannot account for this strong emission, dubbing the high energy emission `hard
excesses', and argue the most likely interpretation is the presence of a Compton
thick (\nh\ $\gtrsim10^{24}$--$10^{25}$ \atpcm), partially covering absorber.

Here we construct disc reflection models and show that they can indeed account
for the strong emission above 10\kev\ in these sources. \S \ref{sec_red}
details our data reduction, while \S \ref{sec_spec}
presents our spectral analysis, and in \S \ref{sec_dis} we discuss our results.

\section{Observations and Data Reduction}
\label{sec_red}

\ngc1365, \1h0419 and \pds456 were observed with \suzaku\ (\citealt{SUZAKU}) in
January 2008, July 2007 and February 2007 respectively. The following sections
detail our data reduction for the X-ray imaging spectrometers (\xis, \citealt{SUZAKU_XIS})
and the hard X-ray detector (\hxd, \citealt{SUZAKU_HXD}).

\subsection{XIS Reduction}

There are four \xis\ detectors on board \suzaku, \xis0, \xis2 and \xis3 are front illuminated,
and \xis1 is back illuminated, however XIS2 experienced a charge leak in November 2006
and has not been in operation since. Using the latest \heasoftv software package we processed
the unfiltered event files for each of the \xis\ CCDs and editing modes
(all observations analysed used both 3x3 and 5x5 editing modes) operational in the
respective observations, following the \suzaku\ Data
Reduction Guide\footnote{http://heasarc.gsfc.nasa.gov/docs/suzaku/analysis/}. We started by
creating new cleaned event files by re-running the \suzaku\ pipeline
with the latest calibration, as well as the associated screening criteria files.
\xselect was used to extract spectral products from these event files, and
responses were generated for each individual spectrum. The spectra and response files for all the front-illuminated instruments operational
were combined using the \ftool\ \addascaspec, and rebinned by a factor of 4. Finally,
spectra were grouped to a minimum
of 20 counts per energy bin with \grppha to allow the use of $\chisq$ minimization
during spectral fitting. In this work, we only make use of the front illuminated
spectra. Quoted uncertainties on model parameters are the 90 per cent confidence intervals
for one parameter of interest, and the uncertainties on count rates are at the
$1\sigma$ level.

\subsection{PIN Reduction}

For the \hxd\ \pin\ detector we again reprocessed the unfiltered event files for the
respective observations following the data reduction guide. Since the \hxd\ is a collimating
rather than an imaging instrument, estimating the background requires individual
consideration of the non X-ray instrumental background (NXB) and cosmic X-ray background
(CXB). The appropriate response and NXB files were downloaded for the respective
observations\footnote{http://www.astro.isas.ac.jp/suzaku/analysis/hxd/}; in each case the
tuned (Model D) background was used. Common good time intervals were obtained with
\mgtime\ which combines the good times of the event and background files, and \xselect\ was
used to extract spectral products. Dead time corrections were applied with \hxddtcor,
and the exposures of the NXB spectra were increased by a factor of ten, as instructed by the
data reduction guide. The contribution from the CXB was simulated using the form of
\cite{Boldt87}, with the appropriate normalisation for the nominal pointing (all observations
were performed with \xis\ nominal pointing), resulting in a CXB rate of $\simeq$0.026\ctps.
The NXB and CXB spectra were then combined using \mathpha\ to give a total background spectrum,
to which a 2\% systematic uncertainty was also added. Finally the data were grouped to have a
minimum of 500 counts per energy bin to improve statistics, and again allow the use of
$\chisq$ minimization during spectral fitting.

This reduction procedure was then verified using the recently released \pin\ reduction
script \hxdpinxbpi\footnote{http://heasarc.nasa.gov/docs/suzaku/analysis/pinbgd.html}.
The two methods were always in excellent agreement.

\begin{figure*}
\begin{center}
\rotatebox{270}{
{\includegraphics[width=120pt]{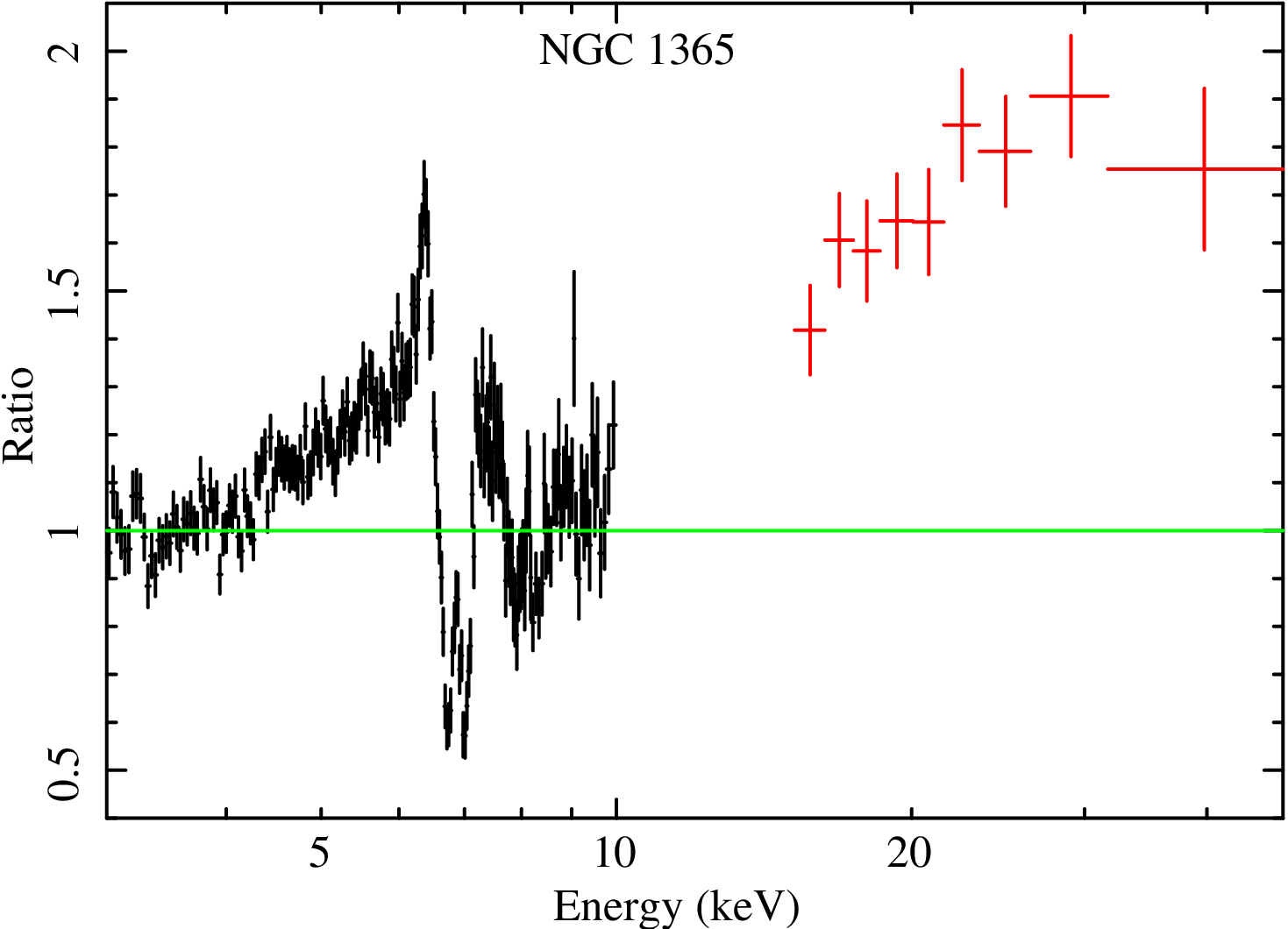}}
}
\rotatebox{270}{
{\includegraphics[width=120pt]{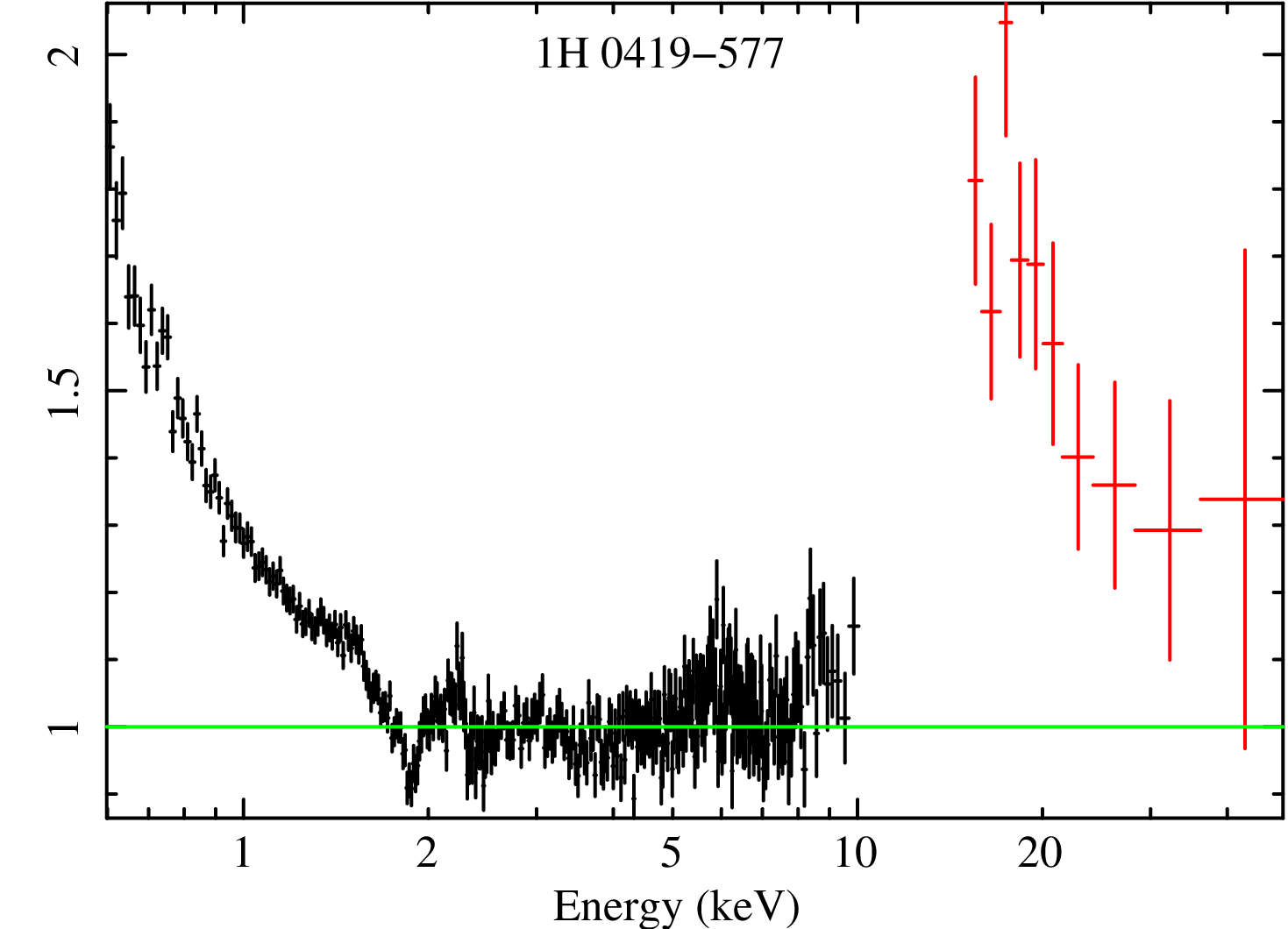}}
}
\rotatebox{270}{
{\includegraphics[width=120pt]{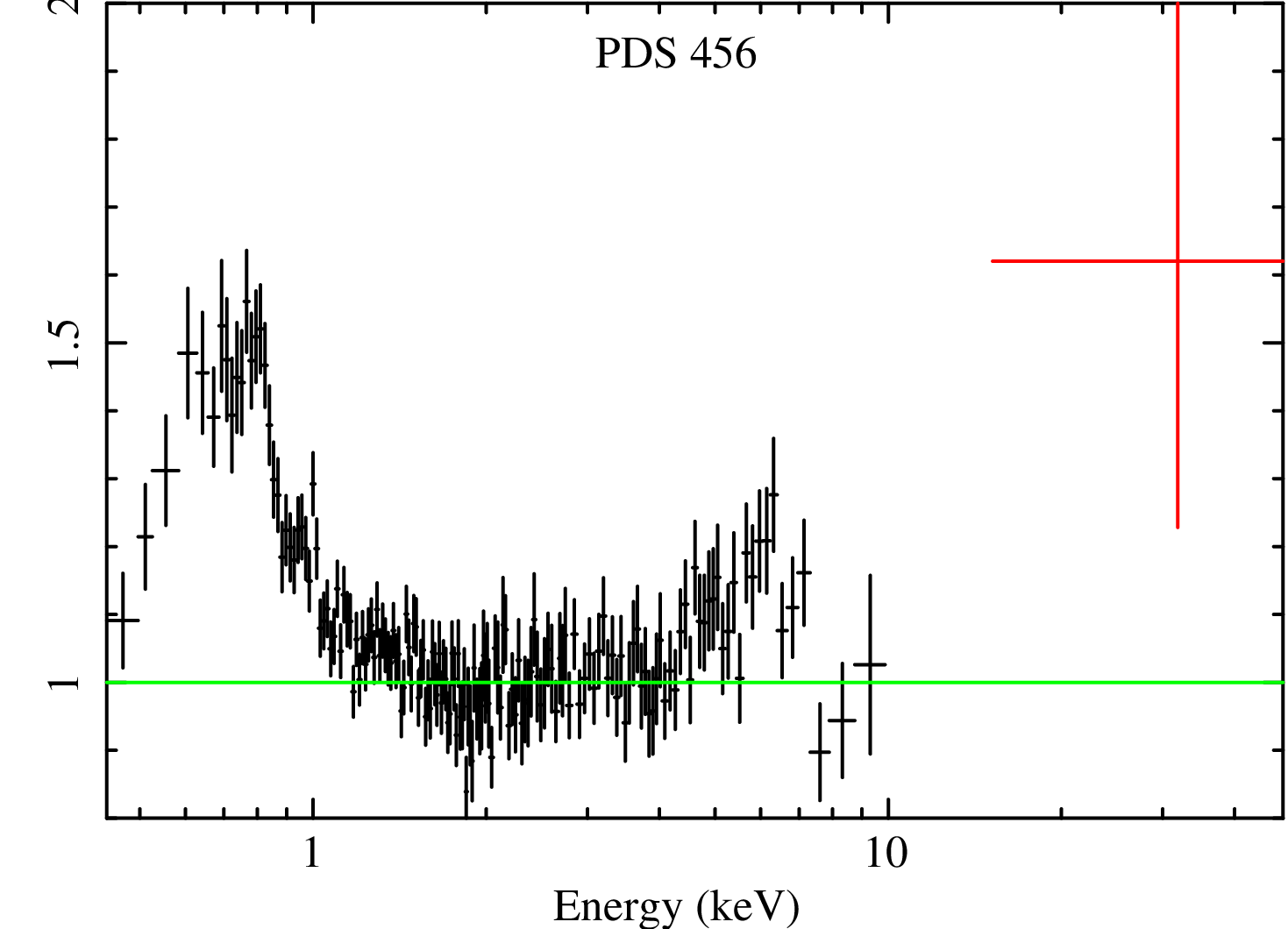}}
}
\end{center}
\caption{Data/model ratio plots of \suzaku\ \xis\ (front illuminated; black) and \pin\ (red) spectra
for (left to right) \ngc1365, \1h0419 and \pds456 to a powerlaw continuum model,
as described in the text. Each of the sources displays strong emission above 10\kev\
which previous authors have claimed cannot be produced using simple reflection models
alone. The data shown have been re-binned for plotting purposes only.}
\label{fig_ratio}
\end{figure*}

\subsection{Specific Source Details}

Here we provide additional details on the data reduction specific to each source.

{\it NGC\thinspace1365}: for this observation, \xis0, \xis1 and \xis3 were all in operation.
Source spectra were extracted from circular
regions of 174'' radius centred on the point source, and background spectra from
another region of the same size, devoid of any obvious contaminating emission, elsewhere on
the same chip. The \pin\ data reduction yielded a source rate of $(11.9 \pm 0.2)\times10^{-2}$\ctps\ 
which is 17.8 per cent of the total observed flux, and the good exposure times
yielded were 161 and 137 \ks\ for each front illuminated \xis\ CCD and the PIN detector respectively.
The reduction performed here is in excellent agreement with that of \cite{Risaliti09c}.

{\it 1H\thinspace0419-577}: here only \xis0 and \xis3 were in operation. In addition to the
usual standard filtering, we excluded \xis\
data with earth elevation angles $<$ 10\deg\ and a cut-off rigidity $>$ 6 GeV in order
to match the reduction procedure adopted in \cite{Turner09b}. Source and background
regions were also circular regions of radius 174'' and were chosen in the same manner
as for \ngc1365. In this case the source was found to contribute 15.0 per cent of the total
observed flux, with the source rate
$(9.4 \pm 0.2)\times10^{-2}$\ctps. The good exposure times
yielded were 179 and 143 \ks\ for each \xis\ CCD and the PIN detector respectively.
Again, our reduction is in good agreement with that of \cite{Turner09b}.

{\it PDS\thinspace456}: \xis0, \xis1 and \xis3 were again in operation for this observation.
Source and background regions were chosen
in the same manner as for \ngc1365, the source region was circular and of radius 174'',
while the background was extracted from various circular regions
scattered around the CCD. The \pin\
detection of \pds456 was much weaker than for the other two sources, only
contributing 2.1 per cent of the observed flux, with a source rate of
$(1.2 \pm 0.2)\times10^{-2}$\ctps\ (considering only the 15.0--50.0\thinspace \kev\
energy range, the observed rate is $(0.6 \pm 0.2)\times10^{-2}$\ctps). The good
exposure times here were 190 and 165 \ks\ for each front illuminated \xis\ CCD and the
PIN detector respectively. The \pin\ rates obtained here are noticeably different from
those presented in \cite{Reeves09}, with their background rate significantly lower. To further test our data reduction in this instance, we also compare the NXB model
with the spectrum obtained during earth occultation (excluding earth elevation angles
$>$ -5\deg), another method for estimating the instrumental background, and
find only a 0.4\% difference between the two. Given our reduction has also been verified
with \hxdpinxbpi, we take our background rate to be the correct one.

\section{Spectral Analysis}
\label{sec_spec}

In order to draw attention to the spectral features present in these sources, we
initially modelled the spectra with a powerlaw continuum, modified by Galactic
absorption, over the 2.0--4.0 and 7.0--10.0\kev\ energy ranges where the intrinsic
continuum may be expected to dominate the observed spectrum in a reflection interpretation. In the case of \ngc1365,
\cite{Wang09} have demonstrated that below $\sim$2\kev\ the X-ray spectrum is
dominated by diffuse thermal emission using the excellent spatial resolution of \chandra.
As such, throughout this work we only study the spectrum above 3\kev\ in this
source, where the contribution from the diffuse emission is negligible. Figure
\ref{fig_ratio} shows the data/model ratios to the powerlaw
continuum for each of the three sources; note that at all times a cross calibration
constant of 1.16 is applied between the \xis\ and \pin\ spectra, as recommended
by the \hxd\ calibration team\footnote{http://www.astro.isas.jaxa.jp/suzaku/doc/suzakumemo/suzakumemo-2008-06.pdf}.
All sources show broad, possibly skewed emission
lines close to $\sim$6.4\kev\ (rest frame), and the ratio spectra of \1h0419
and \pds456 also show excess smooth emission at soft energies ($\lesssim$2\kev).
Both of these are expected from reflection from an accretion disc extending
to within a few gravitational radii (\rg=$GM/c^{2}$) of the central black hole,
where strong gravitational redshifts blur and distort the reflected emission
(see \eg \citealt{Fabian89}, \citealt{Crummy06},
\etc). There are also absorption like features seen at $\sim$2\kev\
in the ratio plots of \1h0419 and \pds456, but these are probably due to uncertainties
in the calibration around the instrumental silicon K edge, so in all further
analysis we exclude 1.7--2.1\kev\ from our modelling.

In addition, it is plain to see that each source displays
fairly strong emission above 10\kev. At these energies, reflection models predict
a broad hump of emission due to the interplay within the reflecting medium between
Compton down-scattering of high energy photons and photoelectric absorption of low
energy photons, a feature commonly referred to as the `Compton hump'. However, 
\cite{Risaliti09c}, \cite{Turner09b} and \cite{Reeves09} have recently claimed
that the high energy emission in these sources cannot be reproduced using simple
reflection models (albeit in the case of \pds456 with a higher \pin\ source rate
than obtained here). In the forthcoming analysis we make use of the self-consistent reflection model \reflionx\ (\citealt{reflion}), which intrinsically
includes iron K-shell absorption and emission, to show that disc reflection
interpretations can in fact successfully model this emission. The key parameters
of \reflionx\ are the iron abundance, $A_{Fe}$,
the photon index of the intrinsic powerlaw continuum, $\Gamma$, and the
ionisation parameter of the reflecting medium, $\xi=L/nR^{2}$. All models include
photoelectric absorption at the Galactic value for each source, and where necessary
we also include a neutral absorption component at the redshift of the respective
galaxies to account for possible additional absorption local to the source.

\subsection{NGC\thinspace1365}
\label{subsec_ngc1365}

We construct a model based around a disc reflection interpretation for the spectrum
of \ngc1365, with the primary components being the intrinsic powerlaw continuum
and a reflected component from a partially ionised accretion disc, modelled with \reflionx.
The effects of strong gravity expected in the innermost regions of the disc are
accounted for by applying the \kdblur\ convolution model, which makes use of the
calculations of \cite{KDBLUR}, to the ionised reflection component. \kdblur\ assumes
a powerlaw form for the emissivity profile of the accretion disc, \ie
$\epsilon(r)\propto r^{-q}$, where the emissivity index $q$ is a free parameter; the
other free parameters are the inner and outer radius and the inclination of the 
accretion disc, \rin, \rout\ and $i$ respectively. In addition
to the broad residuals apparent in Fig. \ref{fig_ratio}, there are also narrow
features present in the spectrum; four absorption lines can be seen between
$\sim$6.7--8.3\kev, and there is also a further narrow iron emission line at $\sim$6.4\kev.
The narrow emission line most likely arises due to reflection from more
distant material, \eg gas in the BLR and/or the dusty torus, so we also include
a neutral reflection component, modelled with \reflionx\ (without \kdblur).
The photon index and iron abundance were required to be the same for the separate
model components, and the redshifts of the reflectors were fixed at that of the host
galaxy ($z=0.0055$; \citealt{RC3}). The absorption lines have been studied in detail by \cite{Risaliti05b}
who identify them with \fexxv\ and \fexxvi\ \ka and \kb absorption. Such
absorption lines are expected to arise due to highly ionised material which would
primarily contribute narrow features, so
initially we simply included four Gaussian absorption lines in order to obtain
estimates for the photon index and iron abundance. Using these, we generated
a photoionisation model for the absorption using XSTAR v2.2.0, in
order to simultaneously account for all four lines with a single, highly ionised absorber.
The free parameters of this absorption model are the column density, ionisation
parameter, iron abundance and outflow velocity of the absorbing medium; for
consistency we require that the iron abundance of the disc and the ionised
absorber are the same.

\begin{figure}
\begin{center}
\rotatebox{270}{
{\includegraphics[width=165pt]{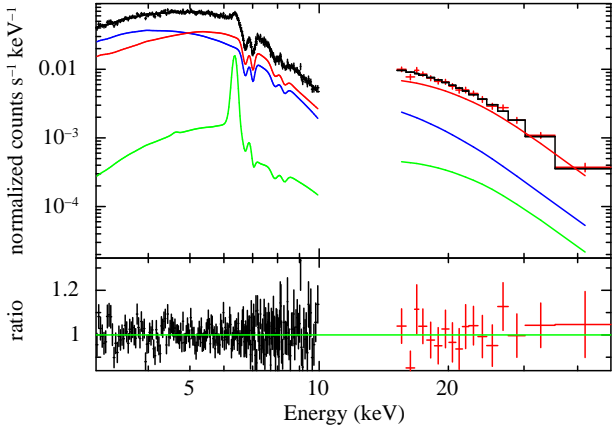}}
}
\end{center}
\caption{The front illuminated \xis\ and \pin\ spectra of \ngc1365,
and the data/model ratios to our best fit reflection models
(see text). The top panel shows the relative contributions of the powerlaw continuum
(blue), disc reflection (red) and neutral reflection (green) components, with the
overall model in black. The data have been re-binned for display purposes only.}
\label{fig_ngc}
\end{figure}

In terms of the physical components included, the model constructed is very similar
to that used in \cite{Risaliti09c} and \cite{Risaliti09a}, the main
difference being the self-consistent treatment of the line and continuum components
in the reflection spectrum provided by \reflionx. During the spectral fitting
we find that the outer radius of the accretion disc is not well constrained, so we
freeze this parameter at 400\rg, the maximal value allowed by \kdblur. This model
provides an excellent fit to the data, with
\rchi = 278/268, parameters are listed in Table \ref{tab_par}, and the fit to
the spectrum is shown in Fig. \ref{fig_ngc}, with the relevant contributions of the
various components highlighted. Notably, the model does not have any problem reproducing
the data above 10\kev.

\subsection{1H\thinspace0419-577}
\label{subsec_1h0419}

Here we also begin with a two component disc reflection model, as for \ngc1365. The photon
index is again required to be the same for all the model
components, the redshift of the reflection is fixed at the value of the
host galaxy ($z=0.104$; \citealt{Thomas98}), and the outer radius of the disc fixed at
400\rg. This model provides a good
representation of the data with \rchi = 423/436, and importantly
fits the \pin\ data well. However, there are still some residual features in the \xis\ 
spectrum. To resolve these we adopt the approach used in \cite{Fabian1h0419} when modelling
\xmm spectra of this source, and allow the emissivity index and ionisation parameter of
the disc to have a radial dependence. Although a crude implementation, splitting
the disc into an inner and outer region, separated by some break radius \rbr,
improves the fit by $\Delta\chisq$ = 18 for 4 extra degrees of freedom. An F-test finds
the probability of such improvement being
coincidental to be only 0.09\%, so we adopt this model as our best fit. Again,
the key parameters are listed in Table \ref{tab_par}, and the
fit to the spectrum is shown in Fig. \ref{fig_1h}.

\subsection{PDS\thinspace456}
\label{subsec_pds456}

\begin{table}
  \caption{Model parameters obtained through modelling the \suzaku\ spectra of \ngc1365,
\1h0419 and \pds456 with a disc reflection interpretation (see text). Parameter
values marked with a * have not been allowed to vary.}
\begin{center}
\begin{tabular}{c c c c}
\hline
\hline
\\[-0.3cm]
& \ngc1365 & \1h0419 & \pds456 \\
\\[-0.3cm]
\hline
\\[-0.3cm]
\nh$_{(G)}$\tmark[a] & 0.134* & 0.2* & $2.3^{+0.3}_{-0.2}$ \\
\\[-0.3cm]
\nh$_{(z)}$\tmark[a] & $95\pm5$ & - & - \\
\\[-0.3cm]
\nh$_{(i)}$\tmark[a] & $113^{+7}_{-11}$ & - & $7^{+5}_{-2}$ \\
\\[-0.3cm]
$\xi_{i}$\tmark[b] & $4830^{+550}_{-800}$ & - & $3140^{+2980}_{-140}$ \\
\\[-0.3cm]
$v_{i}$\tmark[c] & $-3120\pm320$ & - & $-96600^{+3300}_{-5000}$ \\
\\[-0.3cm]
$\Gamma$ & $2.01\pm0.04$ & $2.12\pm0.02$ & $2.35^{+0.05}_{-0.03}$ \\
\\[-0.3cm]
\feabund\tmark[d] & $2.5\pm0.2$ & $1.3\pm0.2$ & $>8.6$ \\
\\[-0.3cm]
$i$ & $57^{+3}_{-2}$ & $53\pm2$ & $70^{+3}_{-2}$ \\
\\[-0.3cm]
\rin\tmark[e] & $1.85\pm0.15$ & $1.4\pm0.1$ & $<2.0$ \\
\\[-0.3cm]
\rout\tmark[e] & $400$* & $400$* & $400$* \\
\\[-0.3cm]
$\xi_{inner}$\tmark[b] & $10^{+32}_{-7}$ & $92^{+50}_{-32}$ & $55^{+10}_{-3}$ \\
\\[-0.3cm]
$\xi_{outer}$\tmark[b] & - & $1.1^{+0.7}_{-0.1}$ & - \\
\\[-0.3cm]
$q_{inner}$ & $5.0^{+1.0}_{-0.5}$ & $>8.7$ & $5.5^{+2.0}_{-1.2}$ \\
\\[-0.3cm]
$q_{outer}$ & - & $5\pm1$ & - \\
\\[-0.3cm]
\rbr\tmark[e] & - & $2.4\pm0.3$ & - \\
\\[-0.3cm]
\hline
\\[-0.3cm]
\rchi & 283/277 & 405/432 & 421/381 \\
\\[-0.25cm]
$F_{0.5-10.0}$\tmark[f] & $12.78^{+0.66}_{-0.83}$ & $32.36^{+0.12}_{-0.13}$ & $6.98^{+0.03}_{-0.04}$ \\
\\[-0.2cm]
$R_{0.5-10.0}$\tmark[g] & $0.23^{+0.16}_{-0.03}$ & $0.67^{+0.23}_{-0.35}$ & $0.15^{+0.06}_{-0.04}$ \\
\\[-0.2cm]
\hline
\hline
\end{tabular}
\label{tab_par}
\end{center}
\small $^a$ Column densities are given in $10^{21}~$\atpcm \\
\small $^b$ Ionisation parameters, given in \ergcmps \\
\small $^c$ Outflow velocities of the ionised absorber, with respect to the host galaxy, given in \kmps \\
\small $^d$ The iron abundances are quoted relative to the solar value \\
\small $^e$ Radii are given in units of the gravitational radius, \rg=$GM/c^{2}$ \\
\small $^f$ Observed 0.5--10\thinspace \kev\ fluxes, given in $10^{-12}$ \ergpcmsqps \\
\small $^g$ 0.5--10\thinspace \kev\ disc reflection fractions \\
\end{table}

Similarly to \1h0419 and \ngc1365, we model the spectrum of \pds456 with a two component
powerlaw plus disc reflection interpretation. As previously stated, we allow for neutral
absorption in excess of the Galactic value, however in this instance \cite{Behar10} find
the excess absorption is located at $z=0$, so we allow the local absorption to vary with
a lower limit of the Galactic value. The photon index is the same for the intrinsic and
reflected components, the redshift of the reflection fixed at at that of the host galaxy,
($z=0.184$; \citealt{Torres97}) and the outer disc radius fixed at 400\rg. In addition,
we also include two absorption lines to model the narrow lines at 7.67 and 8.13\thinspace\kev\
identified by \cite{Reeves09}. As with the absorption lines in NGC\thinspace1365, the
absorbing material is expected to be highly ionised, so we adopted the same procedure and initially
modelled them with Gaussian absorption lines to obtain estimates of the photon index and iron
abundance, then used these to generate a photoionisation absorption model using XSTAR. The free parameters for this
absorption model are the same as that generated for NGC\thinspace1365, and again we require that
the accretion disc and the absorbing medium have the same iron abundance. This reflection-based
model provides a good fit to the observed spectrum, with \rchi = 414/379. As shown in Fig.
\ref{fig_pds}, the high energy data are well modelled, as is the strong emission feature at
$\sim$0.8\thinspace \kev\ (see Fig. \ref{fig_ratio}), identified here with iron L-shell
emission, as in \cite{Reeves09}. Model parameters are listed in Table \ref{tab_par}.

\section{Discussion and Conclusions}
\label{sec_dis}

\begin{figure}
\begin{center}
\rotatebox{270}{
{\includegraphics[width=165pt]{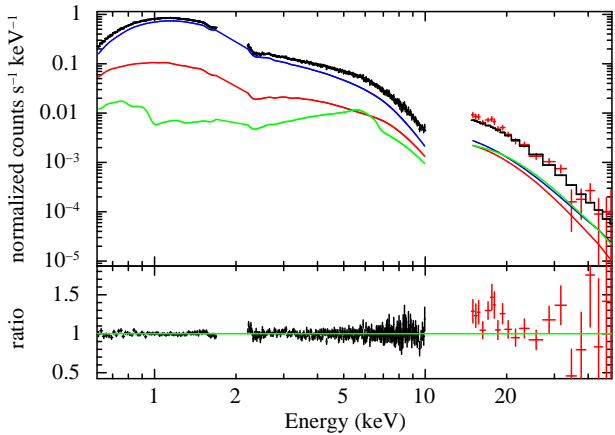}}
}
\end{center}
\caption{As in Fig. \ref{fig_ngc} but now for \1h0419. Here, the red and green components
are the emission from the inner and outer regions of the accretion disc. Again, the data
have been re-binned for display purposes only.}
\label{fig_1h}
\end{figure}

\begin{figure}
\begin{center}
\rotatebox{270}{
{\includegraphics[width=165pt]{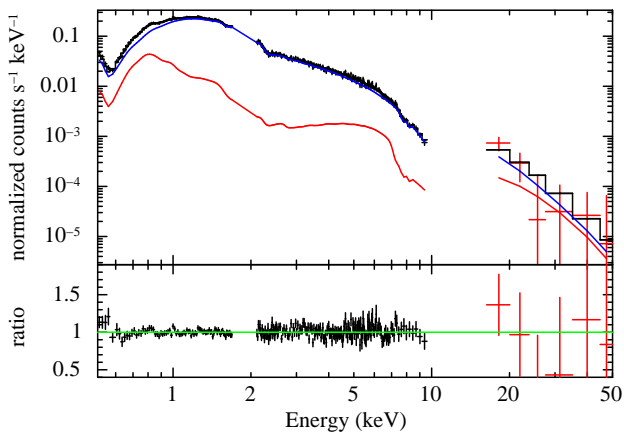}}
}
\end{center}
\caption{As in Figs. \ref{fig_ngc} and \ref{fig_1h}, but now for \pds456.
Again, the data have been rebinned for display purposes only.}
\label{fig_pds}
\end{figure}

With the advent of the \suzaku\ X-ray observatory, and the HXD instrument on board,
observations of sources displaying strong emission above 10\kev\ have come to light.
The most prominent examples to date are \ngc1365, \1h0419 and \pds456, for which
\cite{Risaliti09c}, \cite{Turner09b} and \cite{Reeves09} have respectively claimed
that the high energy emission is too strong to be explained with reflection
models alone. In each case the emission is described as a `hard excess' and
interpreted as observational evidence for the presence of a partially covering,
Compton thick absorber. We have investigated these claims by constructing models
based on disc reflection for each of these sources to test whether they truly
cannot reproduce the observed high energy spectra.

The main conclusion of this work is that for each of the sources analysed, as shown
in Figs. \ref{fig_ngc}, \ref{fig_1h} and \ref{fig_pds}, disc reflection models can successfully reproduce the
observed data above 10\kev, disputing the previous claims to the contrary.
It must be stressed that this does not rule out the presence of Compton thick,
partially covering absorbers in these sources, indeed complex absorption remains a viable
alternative interpretation to the observed spectra; we are demonstrating that disc
reflection can reproduce the broadband data without commenting in depth on the
absorption interpretation. The
reasons we have been able to fit such models to the data where the previous authors
could not differ for each source. In the case of \ngc1365 the model constructed
is physically very similar to that used in \cite{Risaliti09c}, but here the reflection
components are modelled with \reflionx\ which self-consistently includes the reflected
continuum and line features, rather than treating the two separately. In addition, the
\pexrav/\pexriv\ models (\citealt{pexrav}) used in \cite{Risaliti09c} as the reflection continuum components
display a much stronger iron absorption edge at 7.1\kev\ than \reflionx, which may also
contribute to an underprediction at higher energies. For \1h0419 it is not clear how
the modelling procedures differ, but we may have
considered a broader parameter range than \cite{Turner09b}. Finally, in the case of \pds456
we find that the total background count rate obtained in \cite{Reeves09} to be significantly
lower than that obtained here, which has been verified via a variety of methods (see
\S \ref{sec_red}). Such underestimation of the background rate would artificially
inflate the \pin\ spectrum with respect to the \xis\ data.

In their analysis of the short term variability within one of the \xmm observations
of \ngc1365, \cite{Risaliti09a} propose a disc reflection interpretation modified
by variable absorption, perhaps due to eclipsing BLR clouds. Since the model proposed here is very similar,
it is not surprising we obtain some similar results, \eg the iron abundance
is $\sim$3 times solar, and the majority of the emission comes from within a few \rg.
This new model can easily be incorporated into a scenario with such variable absorption,
and a similar time resolved analysis would be useful to disentangle the curvature in the
time averaged spectrum due to the broad iron line and variations
in absorption. Given the spectrum obtained with the long \xmm observation (see \citealt{Risaliti09b}),
the absorption does seem to vary on long time scales, though there may still be
other interpretations for the short term variability. The iron abundance obtained is
in excellent agreement with that found for the ring of star formation $\sim$1.3\thinspace kpc
from the nucleus, but the more distant diffuse emission requires a sub-solar abundance
($\lesssim$0.4 solar; \citealt{Wang09}, \citealt{Guainazzi09}), so it would appear the
metallicity in \ngc1365 has some radial dependence. Although the general
agreement of our model with that of \cite{Risaliti09a} is good, the disc inclination
obtained here is significantly higher, and is closer to that expected for an
obscured AGN in the classical unified theory (\citealt{AGNunimod}).

Using XSTAR, the absorption lines present in NGC\thinspace1365
and PDS\thinspace456 have been successfully modelled as being due to highly
photoionised material. The absorption systems in these two sources have been studied
in detail by \cite{Risaliti05b} and \cite{Reeves09} respectively, including
considerations of the possible origins of the absorbing material. Our modelling
is broadly consistent with the previous work of these authors, and as we are
interested primarily in the continuum we do not embark on any lengthy discussion
of these features. There are some differences in the column densities obtained in
both cases, but these are due to our requirement that the absorbers have the
same iron abundance as the accretion discs; in both cases we find iron to have a
super-solar abundance, while the previous work on these absorbers assumed solar
abundances.

The flux of \1h0419 during the \suzaku\ observation presented here shows the source
to be in a high flux state according to the nomenclature used in \cite{Fabian1h0419},
in which a light-bending interpretation is presented for the spectral variability
observed across a number of \xmm observations. In this interpretation, \1h0419 displays
an anti-correlation between flux and reflection fraction, with
which the low reflection fraction and high flux observed here is consistent. In their
model, \cite{Fabian1h0419} also include an edge component to model \ovii absorption,
and note that the depth of the edge also displays an anti-correlation with flux.
No such absorption is required here, but this is consistent with the observed trend as
the flux is higher than in any \xmm observation. The iron abundance obtained
here is not consistent with the previous work, since this is not expected to vary
with time these observations should be revisited with this in mind. By the same nature, a similar multi-mission analysis to that performed by \cite{Behar10}
is required for the previous observations of \pds456 to test whether the disc
reflection model presented here remains acceptable. The iron abundance obtained
in our analysis of this source is high, which is a consequence of the observed
iron K and L-shell emission features (see Fig. \ref{fig_ratio}). In this respect, \pds456 may be
similar to 1H\thinspace0707-495 (\citealt{FabZog09}, \citealt{Zoghbi2010}).
A brief inspection of the \xmm\ observation taken in the same
year (although not simulatneously) shows a tentative hint of an emission
feature at $\sim$0.8\,\kev\ in the \epicpn\ spectrum, but there are no narrow
features in the \rgs\ spectrum, so if the same feature is present it must
be fairly smooth and broad.

The inclination obtained from our spectral fit to PDS\thinspace456 is higher
than would be expected for an unobscured nuclei in the unified scheme for
AGN, and like the iron abundance is also primarily driven by the detection of iron
L-shell emission. \cite{Lawrence10} present
a geometric model of the inner regions of galactic nuclei which attempts to
reproduce the observed fraction of obscured AGN. The model involves the misalignment
of the inner regions of the accretion disc with the obscuring material, expected to be the
outer regions of the disc and the structures beyond, such as the dusty torus \etc\ 
In the context of PDS\thinspace456, with this model it is possible that the structures
responsible for the obscuration are warped away from the line of sight. The
inclination obtained from modelling the X-ray spectrum is that of the innermost regions
of the accretion disc; if the outer regions are warped away from
our line of sight in this source, it becomes possible to reconcile this high
inclination for the inner disc with the fact that PDS\thinspace456 is an unobscured quasar.
Multi-wavelength studies of a few nearby galaxies in which such substructure could
be resolved do seem to hint that misalignments may be present (\eg NGC\thinspace4151,
NGC\thinspace1068 and Cygnus A; see \S 4.4 in \citealt{Lawrence10} for a summary of the
observations). An alternative explanation comes from the possible evolution in the relative
numbers of type 1 and 2 sources with luminosity seen in X-ray studies of AGN (see \eg
\cite{Hasinger08}, although \cite{Lawrence10} argue that this may not be the case). There
appear to be fewer type 2 sources at higher X-ray luminosities, which suggests that
the opening angle of the obscuring material increases with X-ray luminosity. 
PDS\thinspace456 is one of the brightest low redshift quasars known, and although the observation
analysed here finds the source with the lowest flux to date it is still emitting
at an intrinsic rest frame luminosity of $L_{0.5-10.0}\sim10^{45}$ \ergps. If the proposed
evolution of the absorber geometry with luminosity is real, PDS\thinspace456 could have
a very wide opening angle and may still be seen as an unobscured quasar despite it's high
inclination. There are a number of other AGN which, whilst being unobscured,
also require a high inner disc inclination from spectral fitting, \eg
Ark\thinspace120, NGC\thinspace7213 and NGC\thinspace7469 in the sample of
\cite{Nandra07}, so any model for the inner regions of galactic
nuclei must somehow be able to incorporate such sources.

\section*{ACKNOWLEDGEMENTS}

DJW and RCR acknowledge the financial support provided by STFC, and ACF thanks the
Royal Society. The authors would also like to thank the referee for his/her useful comments.

\bibliographystyle{mnras}

\bibliography{/home/dwalton/papers/references}

\label{lastpage}

\end{document}